\documentclass[12pt]{article}
\pdfoutput=1
\usepackage[english]{babel}
\usepackage[latin1]{inputenc}
\usepackage[normalem]{ulem}

\usepackage[intlimits]{amsmath}
\usepackage{amssymb}
\usepackage{graphicx}
\usepackage[text={16.4cm,23.7cm}]{geometry}

\newcommand{\D}{\text{d}}

\begin{document}

\title{\vspace{-2cm} 
{\normalsize
\flushright TUM-HEP 874/13\\}
\vspace{0.6cm} 
\bf Determination of the Cosmic Antideuteron Flux in a Monte Carlo approach\\[8mm]}

\author{Alejandro Ibarra and Sebastian Wild\\[2mm]
{\normalsize\it Physik-Department T30d, Technische Universit\"at M\"unchen,}\\[-0.05cm]
{\it\normalsize James-Franck-Stra\ss{}e, 85748 Garching, Germany}
}

\maketitle

\begin{abstract}
We investigate in this paper the antideuteron flux produced in high energy collisions of cosmic rays with the interstellar matter. We employ the Monte Carlo generator DPMJET-III together with the coalescence model to simulate, in an event-by-event basis, the antideuteron production in cosmic ray collisions in our Galaxy. Then, we use a diffusion model to calculate the expected flux. We find a secondary antideuteron flux at the Earth which is approximately a factor of two smaller than in previous calculations.
\end{abstract}

\section{Introduction}
\label{sec:introduction}

The existence of cosmic antiprotons produced in high energy collisions of cosmic rays with the interstellar gas was suggested by Fradkin \cite{Fradkin:1955jr} shortly after the discovery of the antiproton at the Bevatron~\cite{Chamberlain:1955ns} and positively observed as a cosmic ray component in the late seventies~\cite{Golden:1979bw,Bogomolov:1979hu}. The measurement of the spectrum was refined in subsequent experiments \cite{Orito:1999re,Matsunaga:1998he,Mitchell:1996bi,Boezio:1997ec,Boezio:2001ac}, culminating with the exquisite recent measurement of the antiproton flux and the antiproton-to-proton fraction by the satellite borne experiment PAMELA \cite{Adriani:2008zq,Adriani:2010rc}. 

The discovery of antinucleons opened the possibility of observing antinuclei in Nature. Indeed, antideuterons were first observed in 1965 in proton-beryllium collisions at the CERN proton-synchrotron~\cite{1965NCimS..39...10M}. It is then natural to expect the observation of antideuterons in cosmic rays, although so far all searches have been fruitless. The best present limit on the cosmic antideuteron flux was set by BESS in the range of kinetic energy per nucleon $0.17\leq T\leq 1.15\,{\rm GeV/n}$, $\Phi_{\bar d}<1.9\times 10^{-4} \,{\rm m}^{-2} {\rm s}^{-1} {\rm sr}^{-1} {\rm (GeV/n)}^{-1}$~\cite{Fuke:2005it}. Interestingly, in the near future, the sensitivity of experiments to the cosmic antideuteron flux will increase significantly, by more than two orders of magnitude. The Alpha Magnetic Spectrometer Experiment (AMS-02) on board of the international space station is currently searching for cosmic antideuterons in two energy windows, $0.2\leq T\leq 0.8~{\rm GeV/n}$ and $2.2\leq T\leq 4.4~{\rm GeV/n}$, with an expected flux sensitivity after five years $\Phi_{\bar d}=1\times 10^{-6} {\rm m}^{-2} {\rm s}^{-1} {\rm sr}^{-1} {\rm (GeV/n)}^{-1}$ in both energy windows~\cite{Doetinchem}. Furthermore, the balloon borne General Antiparticle Spectrometer (GAPS) will undertake, starting in 2014, a series of flights at high altitude in the Antarctica searching for cosmic antideuterons. In the first phase, a long duration balloon (LDB) flight will search for antideuterons in the range of kinetic energy per nucleon $0.1\leq T\leq 0.25~{\rm GeV/n}$, with a sensitivity $\Phi_{\bar d}=1.2\times 10^{-6} {\rm m}^{-2} {\rm s}^{-1} {\rm sr}^{-1} {\rm (GeV/n)}^{-1}$, while in the second, the ultra long duration balloon (ULDB) flight will search in the range $0.1\leq T\leq 0.25~{\rm GeV/n}$, with a sensitivity $\Phi_{\bar d}=3.5\times 10^{-7} {\rm m}^{-2} {\rm s}^{-1} {\rm sr}^{-1} {\rm (GeV/n)}^{-1}$~\cite{Doetinchem}. 

The expected fluxes of cosmic antideuterons produced in collisions of high energy cosmic rays with the interstellar gas have been previously investigated in~\cite{Duperray:2005si,Donato:2008yx}. A precise calculation of the secondary antideuteron flux is important  not only to assess the prospects to observe antideuterons at AMS-02 or GAPS, but also to determine the background fluxes to search for primary antideuterons from exotic sources, such as dark matter annihilations~\cite{Donato:1999gy,Baer:2005tw,Brauninger:2009pe,Kadastik:2009ts,Cui:2010ud,Dal:2012my,Ibarra:2012cc}, dark matter decays~\cite{Ibarra:2009tn,Kadastik:2009ts,Cui:2010ud,Ibarra:2012cc} or black hole evaporation~\cite{Barrau:2002mc}. Unfortunately, the calculation of the secondary fluxes is hindered by the scarce experimental information on the differential cross section for antideuteron production in high energy collisions, hence many approximations and {\it ad hoc} assumptions had to be adopted.

In this work we aim to circumvent some of these assumptions by employing a Monte Carlo generator together with the coalescence model to simulate in an event-by-event analysis the antideuteron production in high energy proton-proton and antiproton-proton collisions. Our method to calculate the source term of secondary antideuterons is described in Section \ref{sec:source-term}. We then calculate in Section \ref{sec:propagation} the antideuteron flux at the Earth using a a diffusion model, including convection, energy loss due to ionization, Coulomb collisions in a ionized plasma and non-annhilating rescatterings on the interstellar medium, as well as diffusive reacceleration caused by interactions with random magnetohydrodynamic waves. Finally, we compare our predicted fluxes to the results of previous literature and to the expected sensitivity of the experiments AMS-02 and GAPS, and we discuss the impact of the various sources of uncertainty in our final result. Our conclusions are presented in Section \ref{sec:conclusions}.

\section{The source term of secondary antideuterons}
\label{sec:source-term}

The source term of secondary antideuterons at the position $\vec r$ with respect to the Milky Way center, $Q^{\text{sec}} \left( T_{\bar{d}},\vec r \right)$, defined as the number of secondary antideuterons produced per unit volume, time and kinetic energy per nucleon $T_{\bar{d}}$, reads ~\cite{Duperray:2005si}
\begin{align}
\label{eqn:Qsec_general}
Q^{\text{sec}} \left( T_{\bar{d}},\vec r \right) =  \sum_{i \in \left\{ p, \, \text{He}, \, \bar{p} \right\}} \sum_{j \in \left\{ p, \, \text{He} \right\}} 4 \pi \, n_j(\vec r) \int_{T_{\text{min}}^{\left( i,j \right)}}^{\infty} \D T_{i} \, \frac{\D \sigma_{i,j} \left( T_i, \, T_{\bar{d}} \right)}{\D T_{\bar{d}}} \, \Phi_{i} \left( T_{i} , \vec r \right) \,.
\end{align}
In this expression, $i$ runs over all relevant incident cosmic rays species, with flux $\Phi_{i} \left( T_{i}, \vec r \right)$ and which are known at the Earth with fairly high accuracy. Besides, $j$ represents the different components of the interstellar medium (ISM) which we assume uniformly distributed over the galactic disk, which extends radially 20 kpc from the center and has a half-thickness $h$ of 100 pc. Lastly,  $\frac{\D \sigma_{i,j} \left( T_i, \, T_{\bar{d}} \right)}{\D T_{\bar{d}}}$ denotes the differential cross section to produce an antideuteron with kinetic energy per nucleon $T_{\bar{d}}$ in a collision of a cosmic ray particle of type $i$ (with kinetic energy per nucleon $T_i$) with an ISM component of type $j$. 

To describe the antideuteron production we will employ the coalescence model~\cite{Butler:1963pp,Schwarzschild:1963zz,Csernai:1986qf,Chardonnet:1997dv,Kadastik:2009ts}, which postulates that the probability of formation of one antideuteron out of an antiproton-antineutron pair with given four-momenta $k_{\bar{p}}^{\mu}$ and $k_{\bar{n}}^{\mu}$ can be approximated as a narrow step function $\Theta \left( \Delta^2+p_0^2 \right)$, where $\Delta^{\mu} = k_{\bar{p}}^{\mu} - k_{\bar{n}}^{\mu}$. In this model, the coalescence momentum $p_0$ is the maximal relative momentum of the two antinucleons that still allows the formation of an antideuteron. One can show~\cite{Kadastik:2009ts} that for $| \vec{k}_{\bar{D}}| \gg p_0$, being $ \vec{k}_{\bar{D}}=\vec k_{\bar{p}}+ \vec k_{\bar{n}}$, this ansatz leads to the following differential antideuteron yield in momentum space:
\begin{align}
\label{eqn:general_coal_formula}
 \gamma_{\bar{d}} \, & \frac{d^3N_{\bar{d}}}{d^3 k_{\bar{d}}} ( \vec{k}_{\bar{d}}) = \frac18 \cdot \frac43 \pi p_0^3 \cdot  \gamma_{\bar{p}} \gamma_{\bar{n}} \frac{d^3 N_{\bar{p}} d^3 N_{\bar{n}}}{d^3 k_{\bar{p}} d^3 k_{\bar{n}}} \left( \frac{\vec{k}_{\bar{d}}}{2}, \frac{\vec{k}_{\bar{d}}}{2} \right) \, .
\end{align}
In absence of a microscopic understanding of the coalescence mechanism, the coalescence momentum $p_0$ should be determined from experiments. It is important to emphasize that the coalescence momentum is not a universal parameter, but it shows a dependence on the underlying process and also on the center of mass energy of the corresponding reaction~\cite{Ibarra:2012cc}. Therefore, the coalescence momentum employed for the antideuteron production in spallations of cosmic rays on the interstellar medium should be extracted from a laboratory experiment of antideuteron production in nucleus-nucleus collisions at the corresponding energy.

It is common in the literature to assume the statistical independence of the antiproton and antineutron production in Eq.(\ref{eqn:general_coal_formula}). In this case the antideuteron yield is proportional to the antiproton yield times the antineutron yield. This assumption is, however, not justified for the production of low energy antideuterons in cosmic ray collisions, which is the energy range of interest for experimental searches, and which takes place very close to the threshold. More specifically, a $p\, p$ collision at $\sqrt{s}=10~{\rm GeV}$ producing one antiproton in the final state must contain at least three protons, reducing the available energy for other particles in the final state to $\sim 6$ GeV or less. The production of an antideuteron requires an additional antineutron and neutron, which is further supressed by the limited phase space available for these particles. This leads to a correlation of $\bar{p}$ and $\bar{n}$ production in collisions close to the antideuteron production threshold $\sqrt{s}_{\text{(thres)}} = 6 m_p$: having already one antinucleon in the final state, the probability of creating another one is reduced. Therefore the factorized coalescence model, which assumes uncorrelated production of the antinuclei, overpredicts the antideuteron yield.

\begin{figure}
\begin{center}
\includegraphics[width=0.8\textwidth]{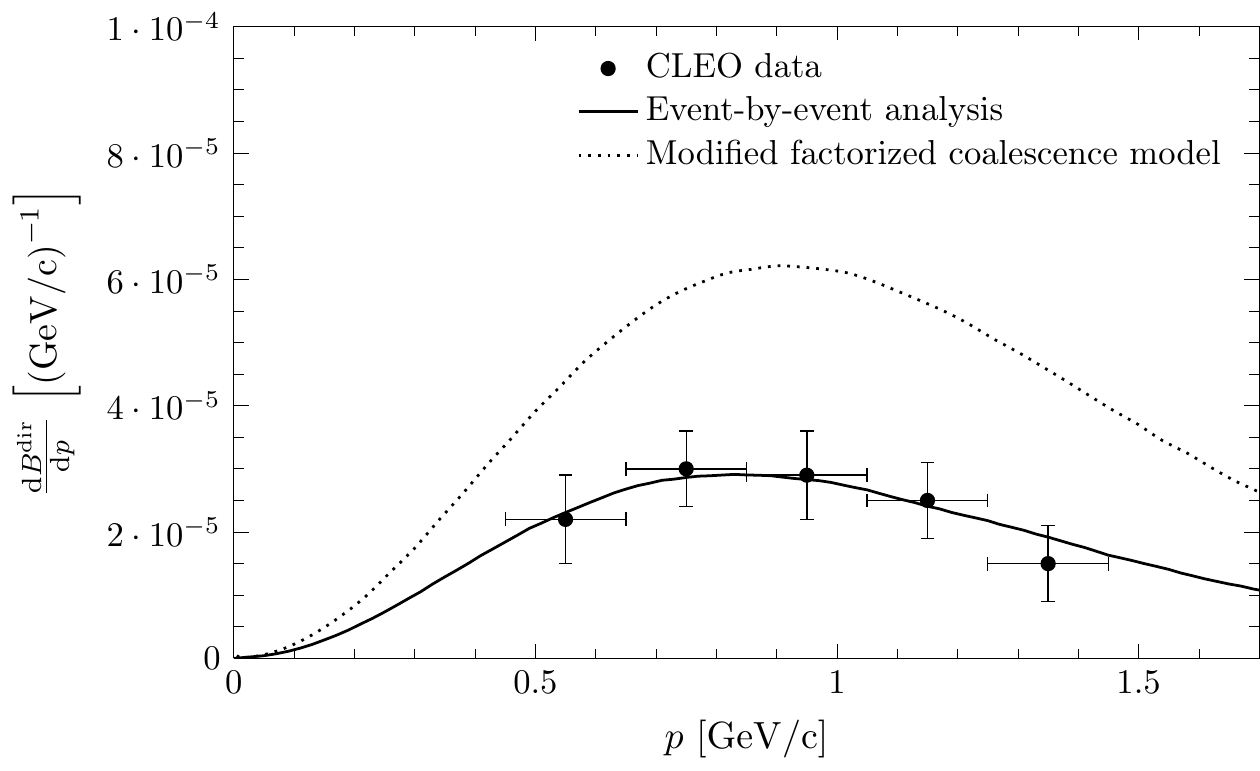}
\caption{Antideuterons from $\Upsilon$ decay. Both calculations are using $p_0=133\,{\rm MeV}$; see the text for details.}
\label{fig:antideuterons_at_upsilon_decay_mfcm}
\end{center}
\end{figure}

This drawback of the factorized coalescence model was circumvented in \cite{Duperray:2005si} by introducing a phase space suppression factor in Eq.(\ref{eqn:general_coal_formula}). Defining the relativistic phase space of $n$ particles with masses $m_i$ as
\begin{align}
 \Phi & \left( \sqrt{s}; m_1,m_2,\dots ,m_n \right) = \nonumber  \int \prod_{i=1}^{n} \left( \frac{1}{\left( 2 \pi \right)^3} \frac{\D^3 p_i}{2 E_i} \right) \delta^3 \left( \sum_{j=1}^{n} \vec{p}_j \right) \delta \left( \sum_{j=1}^{n} E_j - \sqrt{s} \right) \,,
\end{align}
and introducing a phase space suppression factor $R_n \left( x \right)$ for $n$ nucleons with masses $m_p$ through
\begin{align}
\label{eqn:def_R_factor}
R_{n}(x) = \frac{\Phi \left( x; m_p,m_p,\dots ,m_p \right)}{\Phi \left( x; 0,0,\dots ,0 \right)} \, ,
\end{align}
the "modified factorized coalescence model" states that the differential antideuteron yield in momentum space reads:
\begin{align}
\label{eqn:mfcm_formula}
 \gamma_{\bar{d}} \, \frac{\D N_{\bar{d}}}{\D^3 k_{\bar{d}}} \left( \vec{k}_{\bar{d}} \right) = R_n \left( \sqrt{s+m_{\bar{d}}^2-2\sqrt{s} E_{\bar{d}}} \right) \cdot \frac18 \cdot \frac43 \pi p_0^3 \cdot \left[ \gamma_{\bar{p}} \frac{\D N_{\bar{p}}}{\D^3 k_{\bar{p}}} \left( \frac{\vec{k}_{\bar{d}}}{2} \right) \right]^2 \, ,
\end{align}
where $\sqrt{s}$ is the total available center-of-mass energy and $m_{\bar{d}} \left( E_{\bar{d}} \right)$ is the mass (energy) of the antideuteron. The index $n$ of the phase space suppression factor is the minimal number of baryons produced together with the antideuteron and reads $n=4$ for a $p\,p$ collision and  $n=2$ for $\bar{p} p$ collision. The modified factorized coalescence model then takes into account the correlation between the antiproton and the antineutron in the $p\,p$ collision due to the limitied phase space. However, it does not take into account possible correlations in the underlying hard process. The importance of these correlations is manifest in Fig.\ref{fig:antideuterons_at_upsilon_decay_mfcm}, where we compare the spectrum of antideuterons from $\Upsilon \left( \mbox{1S} \right)$ decays predicted by an event-by-event analysis implemented in PYTHIA 8 \cite{Sjostrand:2007gs} for the value  $p_0=133$ MeV (taken from \cite{Ibarra:2012cc}) with the spectrum predicted by the modified factorized coalescence model, calculated using Eq.~(\ref{eqn:mfcm_formula}) with an antiproton spectrum simulated also with PYTHIA 8 and for the same value of the coalescence momentum $p_0=133$ MeV. We also show for comparison the spectrum measured by CLEO~\cite{Asner:2006pw}. As apparent from the plot, taking into account the correlations in the hard process reduces the antideuteron yield by approximately a factor of two. 

\begin{figure}
\begin{center}
\includegraphics[width=0.49\textwidth]{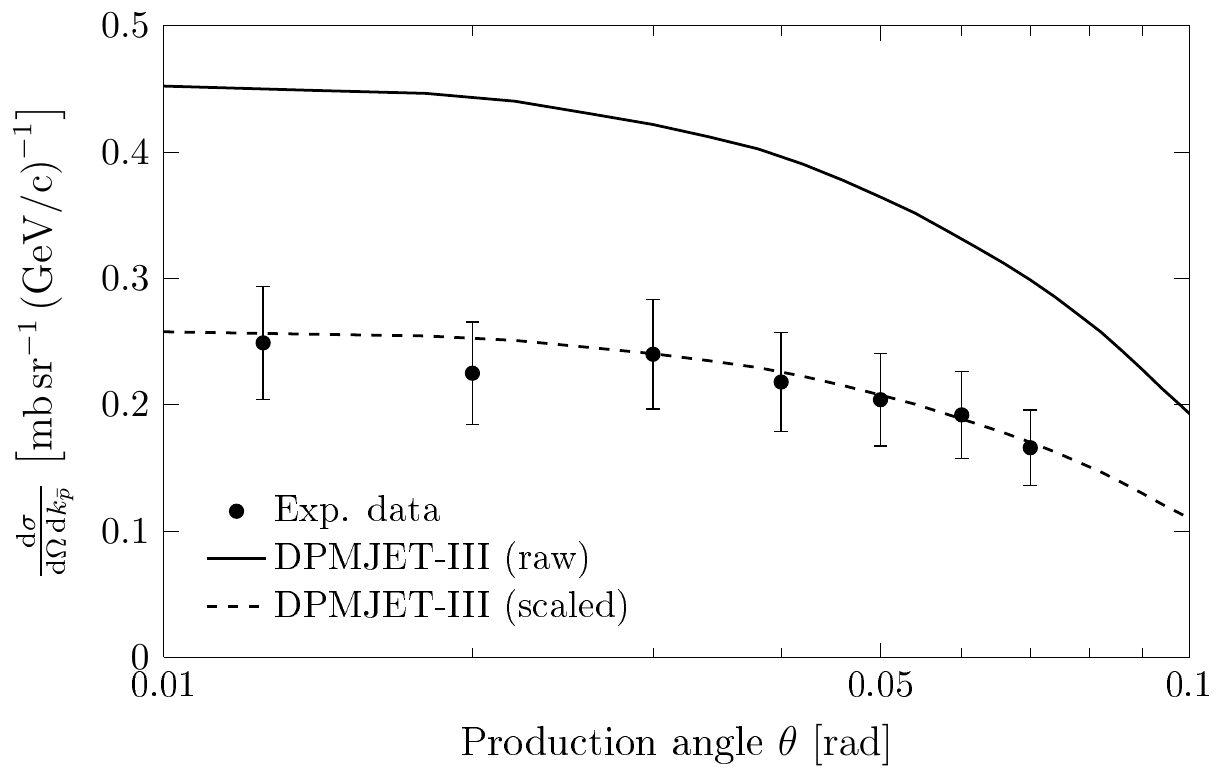}
\includegraphics[width=0.49\textwidth]{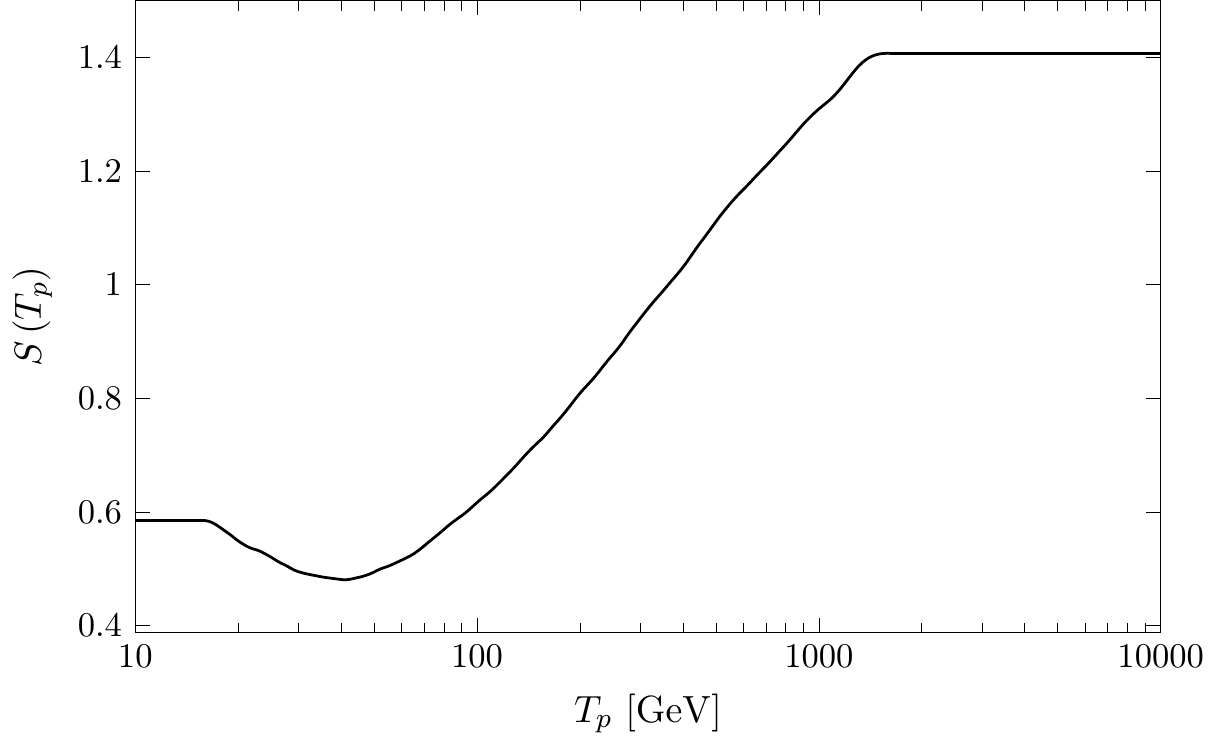}
\caption{{\it Left plot}: Antiproton production cross section in $p\,p$ collisions predicted by DPMJET-III compared to the experimental data from~\cite{Allaby:1970jt}, where $\theta$ is the production angle of the antiproton, while the antiproton momentum $k_{\bar{p}}$ is fixed to 4.5 GeV/c. We also present the corresponding predictions scaled by the function $S \left( T_p \right)$, as defined in the text. {\it Right plot}: Scaling function $S \left( T_p \right)$ as a function of the proton kinetic energy.}
\label{fig:dpmjet}
\end{center}
\end{figure}

In view of this limitation of the modified factorized coalescence model we will employ in this paper a full event-by-event simulation using the event generator DPMJET-III~\cite{Roesler:2000he}, an implementation of the two-component Dual Parton Model that is in particular designed for low-energetic hadron-hadron collisions, such as those relevant for experiments searching for cosmic antideuterons.  Despite the fairly accurate results of DPMJET-III in reproducing the average multiplicity of protons, pions and kaons in $p\,p$ collisions~\cite{Roesler:2001mn}, DPMJET-III overproduces antiprotons in the most relevant region of center of mass energy. This is illustrated in Fig.~\ref{fig:dpmjet}, left plot, which shows the antiproton production differential cross section in $p\,p$ collisions at an incident kinetic energy $T_p=18.3\,{\rm GeV}$ predicted by DPMJET-III compared to experimental data from the S61 experiment at the CERN PS~\cite{Allaby:1970jt}. Evidently, DPMJET-III overestimates the $\bar{p}$ production cross section by a more or less constant factor of $\sim 2$. Therefore we use DPMJET-III in a slightly modified form. First, we extract the total antiproton multiplicity per $p\,p$ collision as a function of the kinetic energy of the impinging proton from DPMJET-III, $n_{\bar{p}}^{\text{DPMJET}} \left( T_p \right)$, and compare it to an interpolating function of experimental values $n_{\bar{p}}^{\text{Exp}} \left( T_p \right)$~\cite{Antinucci:1972ib}. Then we define a scaling factor via
\begin{align}
 S \left( T_p \right) := \frac{n_{\bar{p}}^{\text{Exp}} \left( T_p \right)}{n_{\bar{p}}^{\text{DPMJET}} \left( T_p \right)} \,.
\end{align}

This function parametrizes the overproduction of antiprotons in DPMJET-III and is shown in Fig.~\ref{fig:dpmjet}, right plot. The behaviour for $T_p \gtrsim 1\,{\rm TeV}$, where no experimental information on the antiproton multiplicity is available, is irrelevant due to the small contribution of these processes to the antideuteron source spectrum; for definiteness, we choose $S \left( T_p \right)$ to be constant for $T_p \gtrsim 1\,{\rm TeV}$. The antideuteron production cross section in $p\,p$ collisions is finally obtained via
\begin{align}
\label{eqn:Dpmjet_Scaling_Dbar}
  \frac{\D \sigma_{pp}}{\D T_{\bar{d}}} \left( T_p, T_{\bar{d}} \right)= \left[ S \left( T_p \right) \right]^2 \frac{\D \sigma_{pp}}{\D T_{\bar{d}}} \left( T_p, T_{\bar{d}} \right) \bigg|_{\text{DPMJET (raw)}} \,, 
\end{align}
where the cross section on the right hand side is evaluated implementing the coalescence model, Eq.~(\ref{eqn:general_coal_formula}), in DPMJET-III with a coalescence momentum  $p_0=152\,{\rm MeV}$, determined in \cite{Ibarra:2012cc} from the measurement of the antideuteron production data in $p\,p$ collisions at $\sqrt{s}=53 \,{\rm GeV}$ at the CERN ISR~\cite{Alper:1973my,Henning:1977mt}. The ansatz Eq.~(\ref{eqn:Dpmjet_Scaling_Dbar}) uses the well-motivated assumption that the antineutron and antiproton spectra are equal, and furthermore that the antideuteron yield is proportional to the product of the antiproton and antineutron yield.

In this way, we calculate $\frac{\D \sigma_{pp}}{\D T_{\bar{d}}}$ for 53 logarithmically binned values of the incident kinetic energy $T_p$ in the range between $25\,{\rm GeV}$ and $10\,{\rm TeV}$, which suffices our purposes. These cross sections can finally be inserted in Eq.~(\ref{eqn:Qsec_general}) in order to obtain the $p\,p$ contribution $Q^{\text{sec}}_{pp} \left( T_{\bar{d}} \right)$ to the secondary antideuteron source spectrum. For this, we furthermore use $n_H = 1.0\,{\rm cm}^{-3}$ and take the interstellar proton flux $\Phi_{p}$ measured by the AMS-01 measurement~\cite{Aguilar:2002ad}. 

\begin{figure}
\begin{center}
\includegraphics[width=0.49\textwidth]{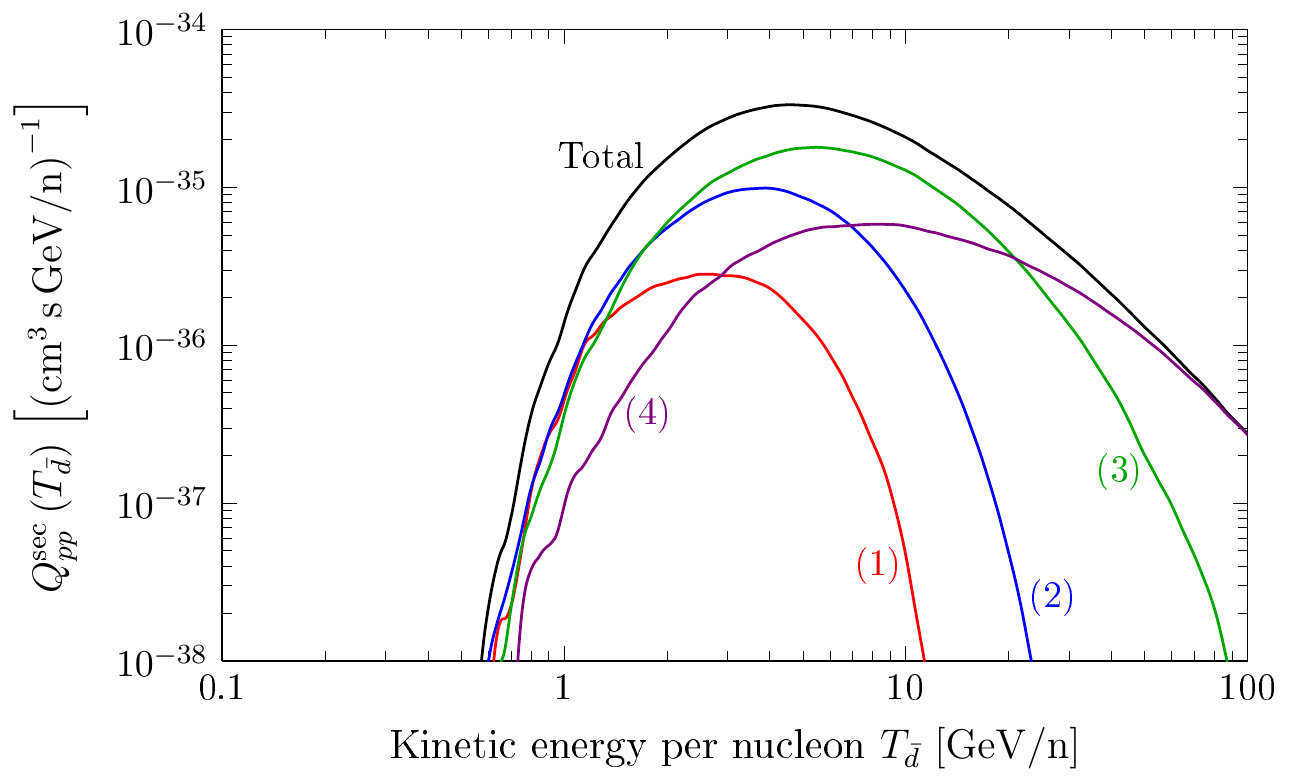} 
\includegraphics[width=0.49\textwidth]{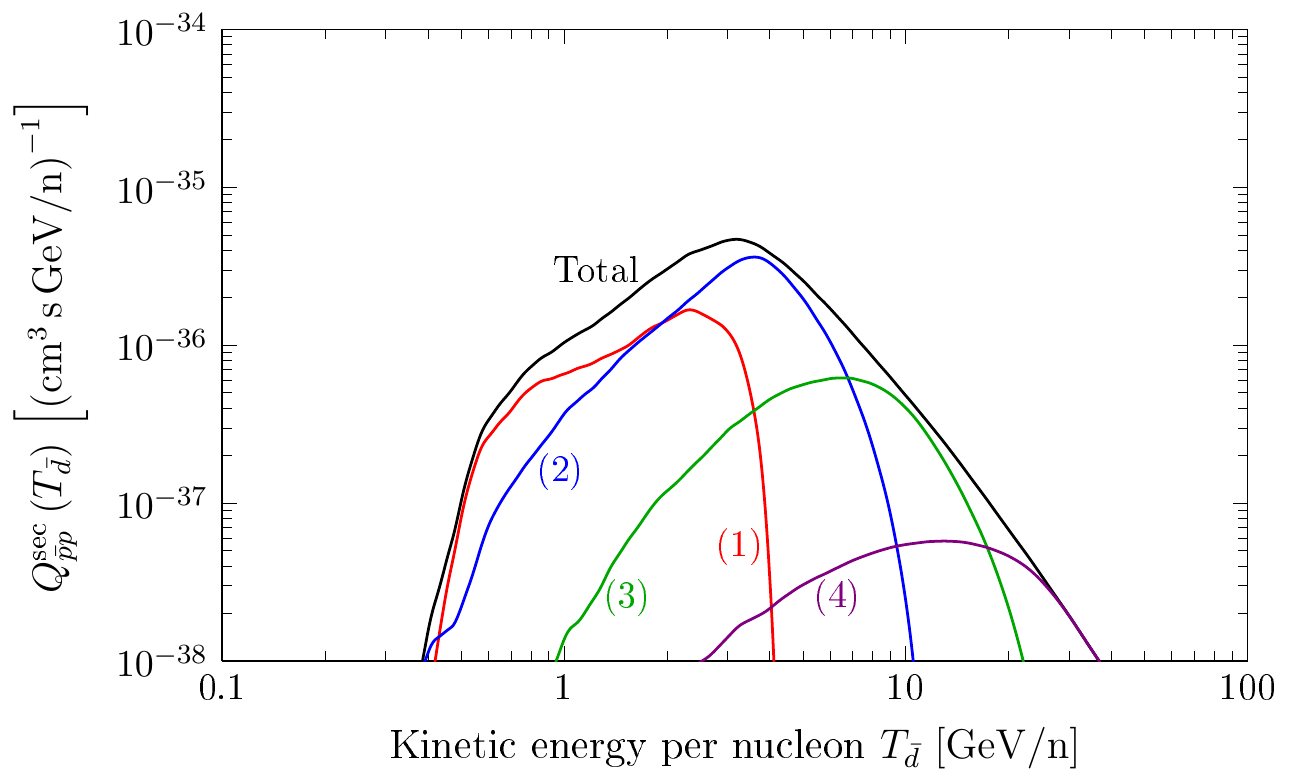} 
\caption{Contributions of different ranges of incident kinetic energies $T_p$ to the antideuteron source spectrum of the $p\,p$ channel (left plot) and the $\bar{p}\,p$ channel (right plot). For the $p\,p$ ($\bar p\,p$) channel, (1) refers to the range $25-42$ GeV ($7-10.6$ GeV), (2) to $42-94$ GeV ($10.6-27$ GeV) , (3) to $94-473$ GeV ($27-67$ GeV) and (4) to $473\,{\rm GeV} - 10\, {\rm TeV}$ ($67-1780$ GeV). The sum of all contributions is shown in black.}
\label{fig:Qsec_contris}
\end{center}
\end{figure}

In Fig.~\ref{fig:Qsec_contris}, left plot,  we show the secondary antideuteron source spectrum $Q^{\text{sec}}_{pp} \left( T_{\bar{d}} \right)$ originating from collisions of cosmic ray protons on the interstellar hydrogen (black curve), together with the different contributions from various ranges of incident proton kinetic energies $T_p$. It is apparent from the plot that the production of antideuterons with $T_{\bar{d}} \lesssim 1\,{\rm GeV/n}$ is highly suppressed, due to the large kinetic energy required for an incident proton to produce an antideuteron,  $T^{\text{thres}}_{p} = 16m_p$. As a result, all produced particles are boosted relative to the center-of-mass frame in the direction of the incident proton, thus being highly unlikely to produce an antideuteron that is almost at rest in the galactic frame. More specifically, one can show that the smallest possible kinetic energy per nucleon of a produced antideuteron is  $T^{\text{min}}_{\bar{d}} = m_p/4 = 0.23\,{\rm GeV/n}$, which requires a proton kinetic energy $T_p \gtrsim 1\,{\rm TeV}$. However, due to the fast decrease of the primary proton flux with the energy, $\propto E^{-2.8}$, the number of low energy antideuterons produced by interactions of high energy protons with the interstellar gas is accordingly very small.

High energy interactions of cosmic ray antiprotons on the interstellar hydrogen and helium can lead, as first noticed in~\cite{Duperray:2005si}, to a non-negligible contribution to the secondary antideuteron source spectrum. In this process, there are only two additional baryons in the final state, hence the threshold for antideuteron production in a $\bar{p}\,p$ reaction is only $T^{\text{thres}}_{\bar{p}} = 6m_p$, while it is $T^{\text{thres}}_p = 16m_p$ for the $p\,p$ process. Therefore, in the energy range relevant for searches for cosmic antideuterons the production cross section in the  $\bar{p}\,p$ channel can be larger than in the  $p\,p$ channel, partially compensating the smaller incident cosmic ray flux of antiprotons in comparison to protons. 

The calculation of the antideuteron yield in the process $\bar{p} \, p \rightarrow \bar{d} \, X$  within the modified coalescence model requires the knowledge of the differential cross sections for $\bar{p}\,p\rightarrow \bar p X$ and  $\bar{p}\,p\rightarrow \bar n X$. Unfortunately the differential cross sections for these two processes are largely unknown, hence previous calculations \cite{Duperray:2005si,Donato:2008yx} assumed
\begin{align}
\label{eqn:Duperray_approx_1}
 E_{\bar{n}} \frac{\D^3 \sigma}{\D k_{\bar{n}}^3} \left( \bar{p} \, p \rightarrow \bar{n} \, X \right) \, \, \simeq \, \, E_{\bar{p}} \frac{\D^3 \sigma}{\D k_{\bar{p}}^3} \left( p \, p \rightarrow \bar{p} \, X \right) \,, \\
\label{eqn:Duperray_approx_2}
  E_{\bar{p}} \frac{\D^3 \sigma}{\D k_{\bar{p}}^3} \left( \bar{p} \, p \rightarrow \bar{p} \, X \right) \, \, \simeq \, \, E_p \frac{\D^3 \sigma}{\D k_p^3} \left( p \, p \rightarrow p \, X \right) \,,
\end{align}
where the differential cross section for $p \, p \rightarrow \bar{p} \, X$ is well known experimentally (see {\it e.g.}~\cite{Tan:1982nc,Duperray:2003bd}). However, the approximation (\ref{eqn:Duperray_approx_1}), when evaluated at the relevant energy range $\sqrt{s} \simeq 10\,{\rm GeV}$ can be questioned from purely kinematical grounds using baryon number conservation: in the reaction on the left hand side of (\ref{eqn:Duperray_approx_1}), at least one additional baryon has to be produced, while the reaction on the right hand side has to have at least three additional baryons in the final state. Therefore one can expect that the latter reaction is suppressed in comparison to the former and hence this approximation used for the evaluation of the antideuteron production is questionable.

\begin{figure}
\begin{center}
\includegraphics[width=0.8\textwidth]{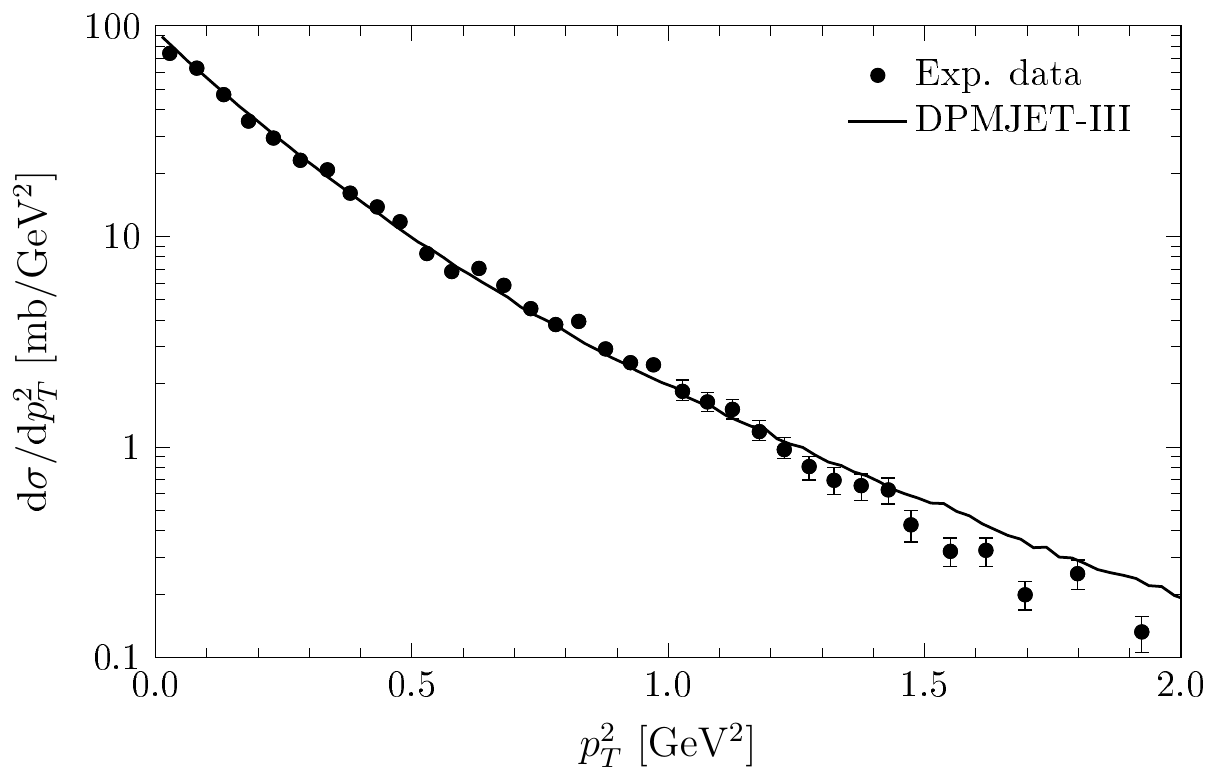}
\caption{Comparison of antiproton production cross section in ${\bar p} p$ collisions in DPMJET-III with experimental data from \cite{Vlasov:1982rm}. }
\label{fig:Vlasov_pbar_p}
\end{center}
\end{figure}

In order to circumvent the limitation of this approach we then simulate the reaction $\bar p  p \rightarrow \bar d X$ using also DPMJET-III on an event-by-event basis. First, we test the validity of DPMJET-III in $\bar p  p$ collisions by comparing the predictions of the (anti)proton yield in the reaction $\bar{p}  p \rightarrow p X$ at $T_{\bar{p}} = 31.1\,{\rm GeV}$ with the measurements at the Mirabelle bubble chamber at IHEP~\cite{Vlasov:1982rm}. As can be seen in Fig.~\ref{fig:Vlasov_pbar_p} the agreement is fairly good (within $\sim 20 \%$) without any rescaling of the (anti)proton yields. Therefore we use DPMJET-III without any modifications for performing the event-by-event analysis of antideuteron production in $\bar{p}\,p$ collisions. In the absence of any information on the coalescence momentum in this process, we use the same value $p_0=152\,{\rm MeV}$ as for the $p\,p$ collisions. Following this procedure, we calculate $\frac{\D \sigma_{\bar{p}\,p}}{\D T_{\bar{d}}}$ for 49 logarithmically binned values of $T_{\bar{p}}$ between $7$ and $1780\,{\rm GeV}$. Lastly, we calculate the secondary source spectrum of antideuterons originated in $\bar p  p$ collisions, $Q^{\rm sec}_{\bar p\,p}(T_{\bar d})$, using Eq.~(\ref{eqn:Qsec_general}) and adopting the values from ~\cite{Bringmann:2006im} for the interstellar antiproton flux $\Phi_{\bar{p}}$. The result is shown in Fig.~\ref{fig:Qsec_contris}, right plot. The antideuteron spectrum extends to slightly lower values of $T_{\bar{d}}$ compared to the $p\,p$ channel, due to the fact that the minimal kinetic energy of the incident particle is only $T^{\text{thres}}_{\bar{p}} = 6m_p$ and therefore the produced particles are less boosted on average. Note that due to the much smaller interstellar antiproton flux $\left( \bar{p} / p \simeq 10^{-4} \right)$, we have only considered incident antiproton energies up to only $\sim 2\,{\rm TeV}$.

Lastly, we also include in our calculation of the source term of secondary antideuterons the contributions from the channels $p$ He, He $p$ and $\bar{p}$ He. We obtain the spectrum of antideuterons produced in a reaction of two nuclei $A$ and $B$ from the corresponding antideuteron spectrum in $p\,p$ collisions by scaling it with an appropriate nuclear enhancement factor $\epsilon_{\text{AB}}^{\bar{d}}$ which, following \cite{1992ApJ394174G}, we assume independent of the energy of the final antideuteron:
\begin{align}
\label{eqn:def_nuclear_enhancement}
 \frac{\D \sigma^{\bar{d}}_{\text{AB}}}{\D T_{\bar{d}}} \left(T_A, T_{\bar{d}} \right) = \epsilon_{\text{AB}}^{\bar{d}} \left( T_A \right) \cdot \frac{\D \sigma^{\bar{d}}_{pp}}{\D T_{\bar{d}}} \left(T_p=T_A, T_{\bar{d}} \right) \,,
\end{align}
$T_A$ being the kinetic energy per nucleon of the nucleus A. We estimate $\epsilon_{\text{AB}}^{\bar{d}}$ for the relevant processes using the ``wounded nucleon'' model \cite{1992ApJ394174G}. With this, the antideuteron production cross sections in the reactions $p$ He, He $p$ and $\bar{p}$ He can be obtained using the corresponding nuclear enhancement factor together with the results of our event-by-event analysis of the $p \,p$ or $\bar{p} \,p$ reaction, respectively. The contributions to the secondary source spectrum are then calculated via Eq.~(\ref{eqn:Qsec_general}), using $n_{\text{He}} = 0.07\,{\rm cm}^{-3}$ and the helium flux measured by AMS-01~\cite{Aguilar:2002ad}. 

\begin{figure}
\begin{center}
\includegraphics[width=0.8\textwidth]{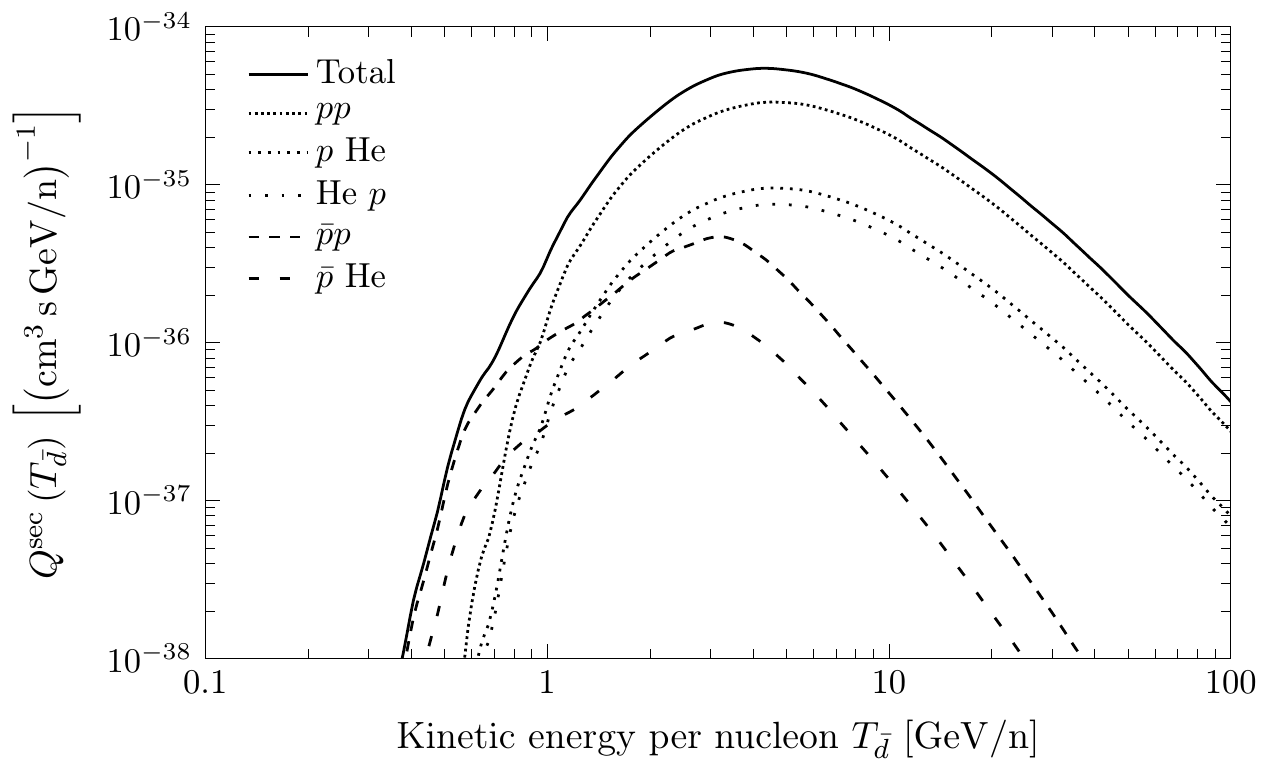}
\caption{Contributions to the secondary antideuteron source spectrum $Q^{\text{sec}} \left( T_{\bar{d}} \right)$.}
\label{fig:QsecContributions}
\end{center}
\end{figure}

The total secondary antideuteron source spectrum $Q^{\text{sec}} \left( T_{\bar{d}} \right)$, {\it i.e.} the sum of the five production channels $p\,p$, $p$\,He, He\,$p$,  $\bar{p}\,p$ and $\bar{p}$\,He is shown in Fig.~\ref{fig:QsecContributions} together with the individual contributions. Evidently, for $T_{\bar{d}} \gtrsim 1\,{\rm GeV/n}$ the most important contribution is the antideuteron production in $p\,p$ collisions, while for $T_{\bar{d}} \lesssim  1 \,{\rm GeV/n}$  the antiproton induced channels dominate. As elaborated above, this can be explained through the different kinematics of the $p\,p$ and $\bar{p}\,p$ reaction, concretely through the different thresholds for antideuteron production in these processes. The sum of the production channels involving helium either as a cosmic ray particle or as a component of the interstellar matter contributes on average $\sim 30 \%$ to $Q^{\text{sec}} \left( T_{\bar{d}} \right)$.

\section{Propagation}
\label{sec:propagation}

Antideuterons produced in collisions of cosmic rays in the interstellar medium propagate in a complicated way before reaching the Earth. The number density of antideuterons in interstellar space, $f \left( E,\vec{r},t \right)$, satisfies the following propagation equation~\cite{Maurin:2001sj,Bringmann:2006im,Berezinskii}:
\begin{align}
\label{eqn:diff_eqn}
 \frac{\partial f \left( E,\vec{r},t \right)}{\partial t} &= \vec{\nabla} \cdot \left( K(E,\vec{r}) \, \vec{\nabla} f - \vec{V}_c f\right) +  2h \, \delta(z) \frac{\partial}{\partial E} D_{EE} (E,\vec{r}) \frac{\partial}{\partial E} f \nonumber  - 2h \, \delta(z) \Gamma_{\text{ann}} f \\[0.1cm]
 & \qquad {} - \frac{\partial}{\partial E} \left\{ 2 \, h \, \delta \left( z \right) \left[ \left(\frac{\D E}{\D t}\right)_{\text{Coul}} + \left(\frac{\D E}{\D t}\right)_{\text{Ion}} \right] f - \frac{E^2-m_{\bar d}^2}{3 \, E} \left( \vec{\nabla} \cdot \vec{V}_c \right) f \right\} \nonumber \\[0.1cm]
 & \qquad {} +  Q^{\rm sec}\left(E,\vec r,t \right) + Q^{\rm ter}\left(E,\vec r,t \right)\;,
\end{align}
where $E$ is the total energy of the antideuteron. We also require the solution $f(E, \vec r, t)$ to vanish at the boundary of the diffusion zone, which is approximated by a cylinder with half-height $L =1-15\,{\rm kpc}$ and radius $R = 20\,{\rm kpc}$. The antideuteron flux $\Phi$ can then be derived from the number density using
\begin{align}
\Phi(E,\vec{r},t) = \frac{v(E)}{4 \pi} f \left( E,\vec{r},t \right) \, ,
\end{align}
with $v(E)$ being the antideuteron velocity. 

We assume in the following that the number density of antideuterons is stationary,  $\frac{\partial f}{\partial t} = 0$, since the typical time scale of antideuteron propagation is small compared to the time scale on which galactic propagation conditions change ~\cite{Blum:2010nx}. The first term in the right-hand side of the propagation equation accounts for the diffusion of charged particles in the Milky Way as a result of scattering on inhomogeneities of the galactic magnetic fields, leading to a random walk of the particles. In our calculation we assume that the corresponding diffusion coefficient is constant in the whole diffusion zone\footnote{For a discussion of the effects of a position dependent diffusion coefficient see, {\it e.g.}~\cite{Perelstein:2010gt}.} and is parametrized as~\cite{Strong:2007nh,Maurin:2001sj} $K(E)= K_0 \, \beta \, \mathcal{R}^{\delta}$, where $\beta=v/c$ and $\mathcal{R}$ is the rigidity, defined as the momentum per unit charge. The second term describes the convective transport induced by a galactic wind of charged particles which we model as $\vec{V}_c = V_c \, \text{sgn}(z) \, \hat{e}_z$~\cite{Maurin:2001sj}. The third term accounts for the diffusive reacceleration in the galactic disk caused by interactions with random magnetohydrodynamic waves, commonly taken into account by diffusion in energy space. Following~\cite{Maurin:2001sj,Donato:2001ms}, we use a spatially constant energy diffusion coefficient $D_{EE}(E)$ which is related to the spatial diffusion coefficient $K(E)$ by
\begin{align}
\label{eqn:definition_of_energy_diffusion_term}
D_{EE}(E) = \frac{2}{9} V_{\text{A}}^2 \frac{E^2 \, \beta^4}{K(E)} \,,
\end{align}
being $V_{\text{A}}$ the  Alfv\'{e}n velocity. The fourth term describes particle losses due to annihilations of antideuterons with protons or helium nuclei in the interstellar gas in the galactic disk with a rate $\Gamma_{\text{ann}}=(n_\text{H}+4^{2/3} n_{\text{He}}) \, \sigma_{\bar{d}p}^{\text{ann}} \, v$, being $n_{\text{H}}=1\,{\rm cm}^{-3}$ and $n_{\text{He}}=0.07\,{\rm cm}^{-3}$. Here it is assumed that the annihilation cross section between an antideuteron and a helium nucleus is related to the annihilation cross section between an antideuteron and a proton by a simple geometrical factor $4^{2/3}$. The annihilation cross section of an antideuteron with a proton is obtained substracting the non-annihilating rescattering (NAR) cross section from the inelastic cross section: $\sigma^{\text{ann}}_{\bar{d} \, p} = \sigma^{\text{inel}}_{\bar{d} \, p} - \sigma^{\text{NAR}}_{\bar{d} \, p}$. The inelastic $\bar{d} p$ cross section can be estimated through $\sigma^{\text{inel}}_{\bar{d} \, p} \left( T_{\bar{d}} \right) \simeq 2 \, \sigma^{\text{inel}}_{\bar{p} \, p} \left( T_{\bar{p}} = T_{\bar{d}} \right)$~\cite{Donato:2008yx}, using the inelastic $\bar{p} p$ cross section from~\cite{Tan:1983de}:
\begin{align}
\label{eqn:cross_section_parametrizations}
\sigma^{\rm inel}_{{\bar p}\,p} \left( T_{\bar{p}} \right) &= 24.7 \, \left( 1 + 0.584 \, T_{\bar{p}}^{-0.115} + 0.856 \, T_{\bar{p}}^{-0.566} \right)\;.
\end{align}
 Besides, for the non-annihilating rescattering cross section, $\sigma^{\text{NAR}}_{\bar{d} \, p} \left( T_{\bar{d}} \right)$, we use the estimation presented in ~\cite{Donato:2008yx} based on experimental data on the total cross section for the charge symmetric reaction $\bar{p} + d \, \rightarrow \, \bar{p} / \bar{n} + d + (\text{n} \pi)$. The fifth term  describes continuous energy losses in the galactic disk, originating from Coulomb collisions in an ionized plasma and from ionization losses, where a part of the energy of the antideuteron is transferred to an electron of an interstellar matter atom or molecule. For the energy loss rates $\left( \frac{\D E}{\D t} \right)_{\text{Coul}}$ and $\left( \frac{\D E}{\D t} \right)_{\text{Ion}}$ we refer to the formulas presented in~\cite{Strong:1998pw}. The sixth term describes adiabatic energy gains or losses which result from non-uniform convection velocities~\cite{Longair:1992ze}, which, adopting $\vec{V}_c = V_c \, \text{sgn}(z) \, \hat{e}_z$, is also proportional to $\delta \left( z \right)$ and hence confined to the Galactic disk. The last two terms account, respectively, for the antideuteron production in high energy interactions of cosmic rays in the galactic disk, being the source term of secondary antideuterons described in Section \ref{sec:source-term} , and for the antideuteron energy loss through non-annihilating rescattering on the interstellar matter in the galactic disk, {\it i.e.}~reactions of the type $\bar{d} + p \, \rightarrow \, \bar{d} + X$ or $\bar{d} + \text{He} \, \rightarrow \, \bar{d} + X$. This process, dubbed tertiary source term, does not change the total number of antideuterons but leads to a redistribution of the antideuteron number density towards lower energies. Formally, the tertiary source term reads~\cite{Duperray:2005si}:
\begin{align}
\label{eqn:tertiary_comp}
Q^{\text{ter}} \left( T_{\bar{d}},\vec r \right) &= 4 \pi n_H (\vec r)\Biggl[ \int_{T_{\bar{d}}}^{\infty} \frac{\D \sigma^{\text{NAR}}_{\bar{d} \, p} \left( \bar{d}(T_{\bar{d}}') + p \rightarrow \bar{d}(T_{\bar{d}}) + X \right)}{\D T_{\bar{d}}} \, \Phi_{\bar{d}} \left(T_{\bar{d}}',\vec r \right) \, \D T_{\bar{d}}' \nonumber \\
 &\qquad {} - \sigma^{\text{NAR}}_{\bar{d} \, p}
 \left( \bar{d} \left( T_{\bar{d}} \right) + p \rightarrow \bar{d} + X \right) \, \Phi_{\bar{d}} \left(T_{\bar{d}} ,\vec r \right) \Biggr] \,,
\end{align} 
where $\Phi_{\bar{d}} \left( T_{\bar{d}} \right)$ is the antideuteron flux at kinetic energy per nucleon $T_{\bar{d}}$ and $\sigma^{\text{NAR}}_{\bar{d} \, p} \left( T_{\bar{d}} \right)$ is the non-annihilating rescattering cross section, introduced after Eq.~(\ref{eqn:cross_section_parametrizations}). For the NAR differential cross section there is unfortunately no experimental information, hence we calculate it following the approach pursued in~\cite{Duperray:2005si,Donato:2008yx} which is based on the measurement by Anderson {\it et al.} of the inclusive $p+p \rightarrow p+X$ cross section at 10, 20 and 30 GeV/c incident proton momentum~\cite{Anderson:1967zzc}. The tertiary contribution coming from the NAR process $\bar{d}+\text{He} \rightarrow \bar{d}+X$ is also included by assuming that all relevant cross sections can be obtained from the corresponding process involving a proton by multiplication by a geometrical factor $4^{2/3}$~\cite{Donato:2008yx}.  Notice that the tertiary source term depends itself on the antideuteron flux, hence the propagation equation is an integro-differential equation which we solve following the method presented in \cite{Donato:2001ms}.

In the propagation equation there are five undetermined parameters: the half-height of the diffusion zone $L$, the normalization of the diffusion coefficient $K_0$, the spectral index $\delta$ of the rigidity in the diffusion coefficient, the convection velocity $V_c$ and the Alfv\'{e}n velocity $V_{\text{A}}$ which have to be determined from observations. We list in Table \ref{tab:MinMedMax} three sets of parameters derived in \cite{Donato:2003xg} compatible with the measurements of the Boron-to-Carbon ration and which yield the the minimal (MIN), medium (MED) and maximal (MAX) primary antiproton flux. The dependence of the final flux on the chosen set of parameters can therefore be considered as the uncertainty band connected to the imprecise knowledge of the propagation parameters.

\begin{table}
 \centering
 \begin{tabular}{|c|c c c c c|} 
  \hline
  Model & $\delta$ & $K_0$ (kpc$^2$/Myr) & $L$ (kpc) & $V_c$ (km/s) & $V_\text{A}$ (km/s) \\ \hline
  MIN & 0.85 & 0.0016 & 1 & 13.5 & 22.4\\
  MED & 0.70 & 0.0112 & 4 & 12 & 52.9\\
  MAX & 0.46 & 0.0765 & 15 & 5 & 117.6\\ \hline
 \end{tabular}
\caption{Definition of MIN, MED and MAX propagation parameters (see text). The values are taken from~\cite{Donato:2003xg}.}
\label{tab:MinMedMax}
\end{table}

Lastly, in order to calculate the flux of a charged cosmic ray species at Earth, one has to take into account the effect of the solar wind. We adopt in this paper the effective theory of a spherically symmetric and charge-independent force field presented in~\cite{1967ApJ149L115G,1968ApJ1541011G,1987AA184119P}. In that simplified model, the net effect of the solar wind is an electric potential generated by the Sun leading to an energy loss of each charged particle. Then the flux at the top of the Earth's atmosphere $\Phi^{\text{TOA}}$ is related to the interstellar flux $\Phi^{\text{IS}}$ through
\begin{align}
\label{eqn:solarModulationFormula}
 \Phi_{\text{TOA}} \left( T_{\text{TOA}} \right) = \frac{p_\text{TOA}^2}{p_\text{IS}^2} \Phi_{\text{IS}} \left( T_{\text{IS}} \right) = \left( \frac{2 m \, \text{A} \, T_{\text{TOA}}+(\text{A} \, T_{\text{TOA}})^2}{2 m \, \text{A} \, T_{\text{IS}}+(\text{A} \, T_{\text{IS}})^2} \right) \Phi_{\text{IS}} \left( T_{\text{IS}} \right)
\end{align}
for a cosmic ray nucleus with mass $m$ and mass number A. As usual, $T$ denotes the kinetic energy per nucleon, i.e.~$T=\frac{E-m}{\text{A}}$. The top of atmosphere and interstellar kinetic energy per nucleon are related via $T_{\text{IS}}=T_{\text{TOA}}+\frac{\left| \text{Z} \right|}{\text{A}} \phi_\text{F}$, where $Z$ is the electric charge of the nucleus in units of $e$ and $\phi_\text{F}$ is the Fisk potential. The latter quantity is the effective parameter controlling solar modulation and varies between $500\,{\rm MeV}$ and $1.3\,{\rm GeV}$ over the eleven year solar cycle. In our calculations we will adopt the value $\phi_\text{F} = 500\,{\rm MeV}$ (corresponding to solar minimum) and we will discuss later the impact on our results of using a different value of the Fisk potential.

\begin{figure}
\begin{center}
\includegraphics[width=0.8\textwidth]{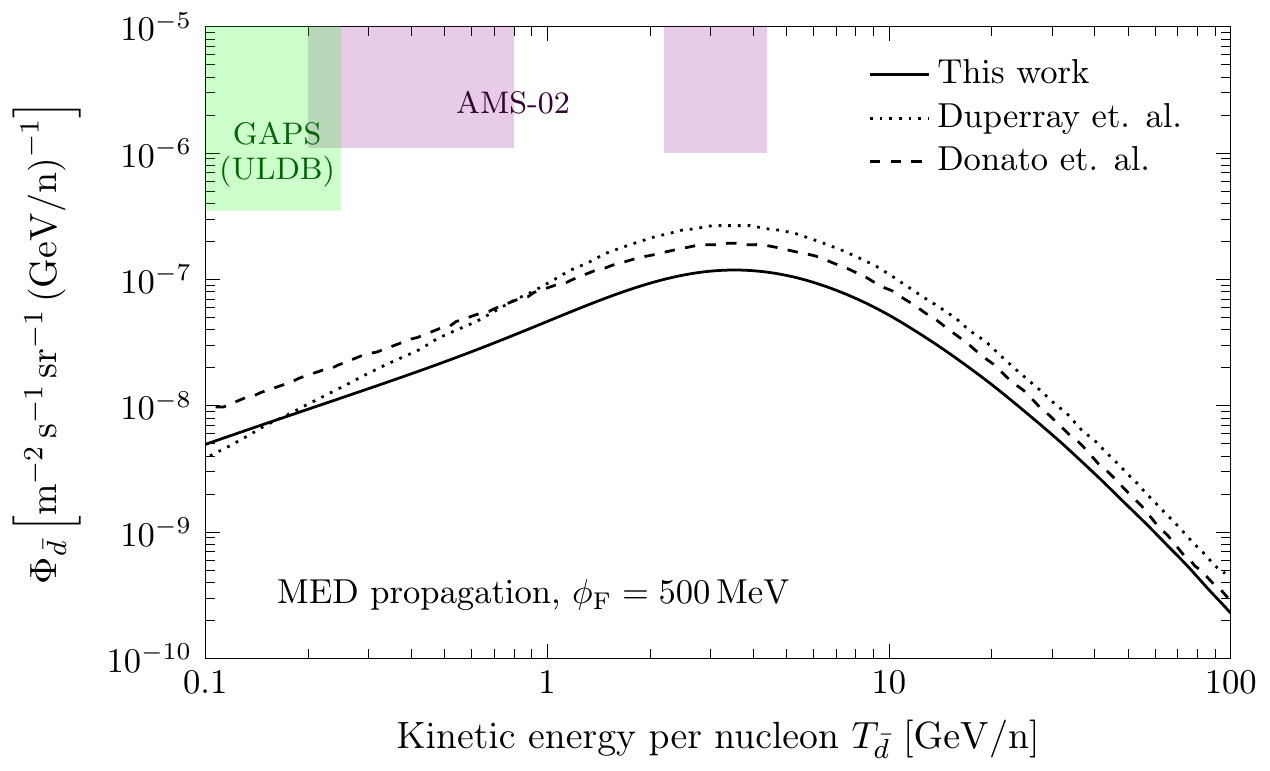}
\caption{Antideuteron background flux at the top of atmosphere (TOA) for the MED propagation model and a fisk potential $\phi_F=500\,{\rm MeV}$. The solid line corresponds to the result of this work, while the dotted and dashed lines are taken from Duperray {\it et al.}~\cite{Duperray:2005si} and from Donato {\it et al.}~\cite{Donato:2008yx}, respectively. The shaded regions indicate the prospected sensitivities of AMS-02 and GAPS ULDB.}
\label{fig:Background_Final}
\end{center}
\end{figure}

\begin{figure}
\begin{center}
\includegraphics[width=0.8\textwidth]{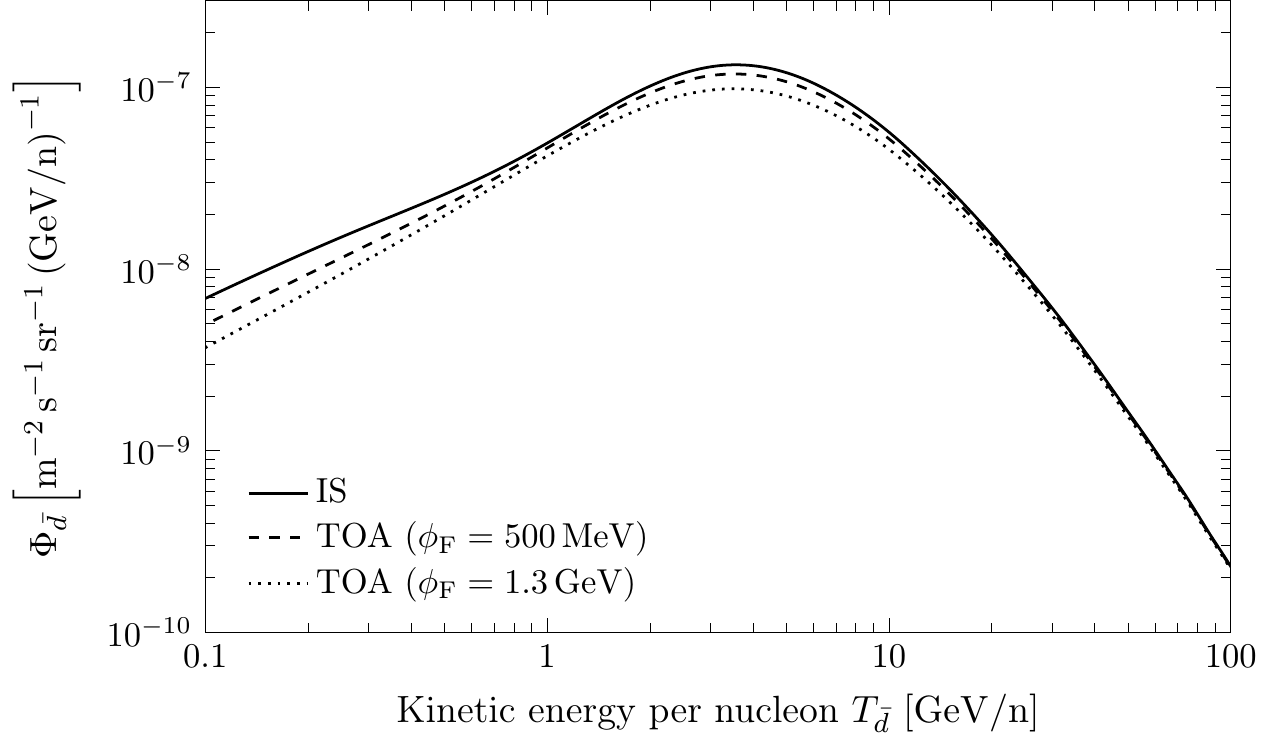}
\caption{Antideuteron background flux in interstellar space (IS) and at the top of the Earth's atmosphere (TOA), for two different values of the Fisk potential $\phi_\text{F}$. All curves assume MED propagation parameters.}
\label{fig:SolarModulation}
\end{center}
\end{figure}

We show in Fig.~\ref{fig:Background_Final} our final result for the antideuteron secondary flux at the top of the Earth's atmosphere (TOA), using a Fisk potential $\phi_\text{F} = 500\,{\rm MeV}$ and the MED propagation model. Note that the antideuteron background flux extends to arbitrary small kinetic energies, in contrast to the secondary source spectrum calculated in Section \ref{sec:source-term} and shown Fig.~\ref{fig:QsecContributions}, which presents a hard cut-off at $T_{\bar d} \sim 0.3$ GeV/n. This is due to the fact that for very low kinetic energies the background flux is dominated by the adiabatic energy loss and the tertiary effect, both significantly redistributing antideuterons from intermediate ($T_{\bar d} \sim 5$ GeV/n) to low ($T_{\bar d} \sim 0.1 - 1$ GeV/n) kinetic energies. We also show, for comparison, the results for the TOA antideuteron secondary flux obtained in the most recent calculations~\cite{Duperray:2005si,Donato:2008yx}, which use the modified factorized coalescence model instead of a full event-by-event simulation, which results in different antideuteron production cross sections in the high-energy reactions. Lastly, we also show in the figure the prospected sensitivities of AMS-02 and GAPS ULDB. Both experiments are clearly not sensitive enough to detect antideuterons from collisions of cosmic rays in the interstellar matter. Translated into number of events, one expects 0.13 events at AMS-02 and 0.024 events at GAPS ULDB. Therefore, the observation of antideuteron events at AMS-02 or GAPS can be interpreted as a signal for an exotic source of antideuterons, although, as argued in \cite{Ibarra:2012cc}, the stringent limits on the primary antiproton flux from the PAMELA measurements of the $\bar p/p$ ratio make this possibility also unlikely. The searches for cosmic antideuterons away for the solar minimum are more pesimistic, due to the larger solar modulation effects. This is illustrated in Fig.~\ref{fig:SolarModulation}, where we show the top of the atmosphere antideuteron flux for values of the Fisk potential $\phi_F=500$ MeV and $\phi_F=1.3$ GeV, as well as the interstellar flux.

\begin{figure}
\begin{center}
\includegraphics[width=0.49\textwidth]{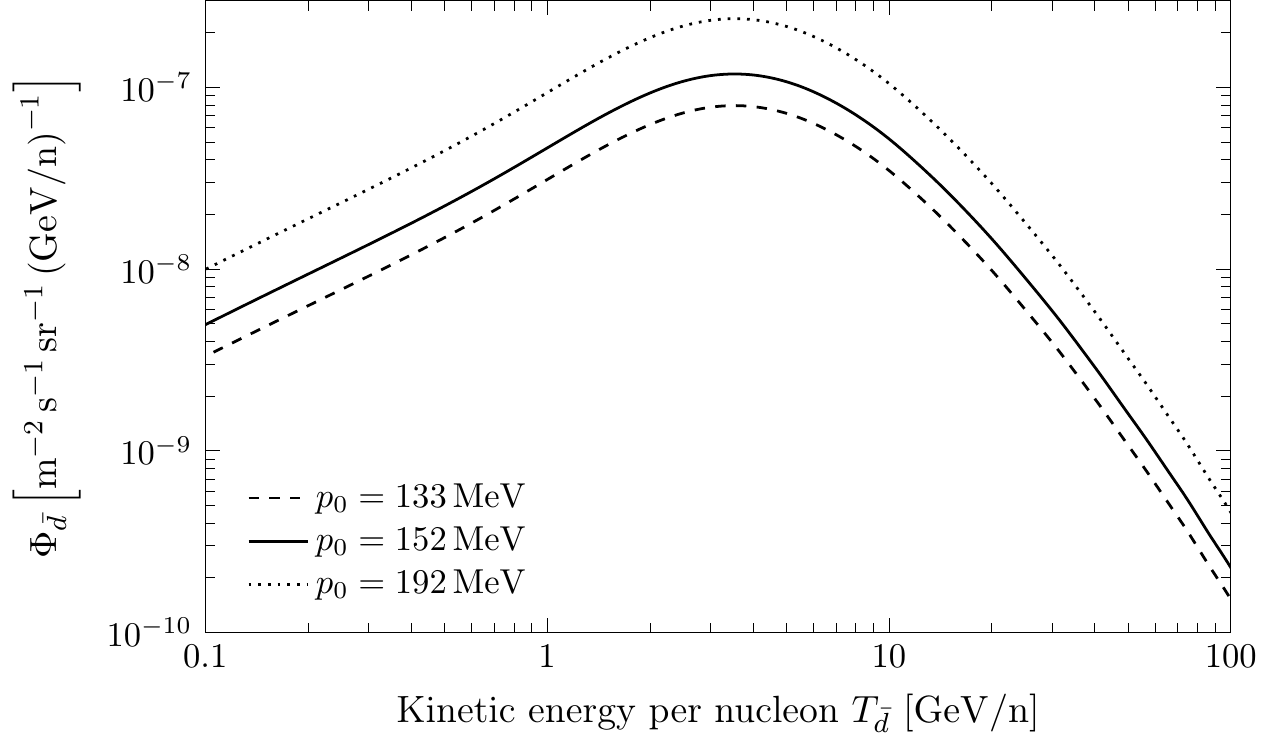}
\includegraphics[width=0.49\textwidth]{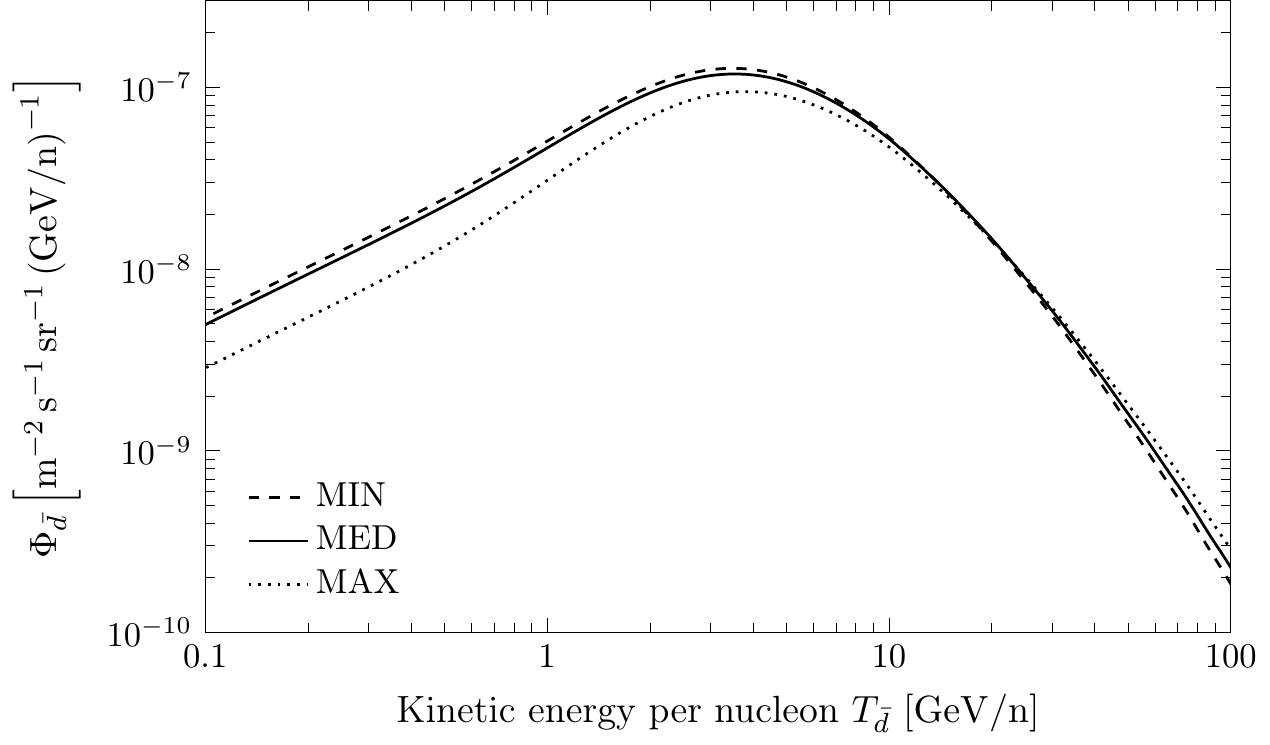} \\\vspace{0.4cm}
\includegraphics[width=0.49\textwidth]{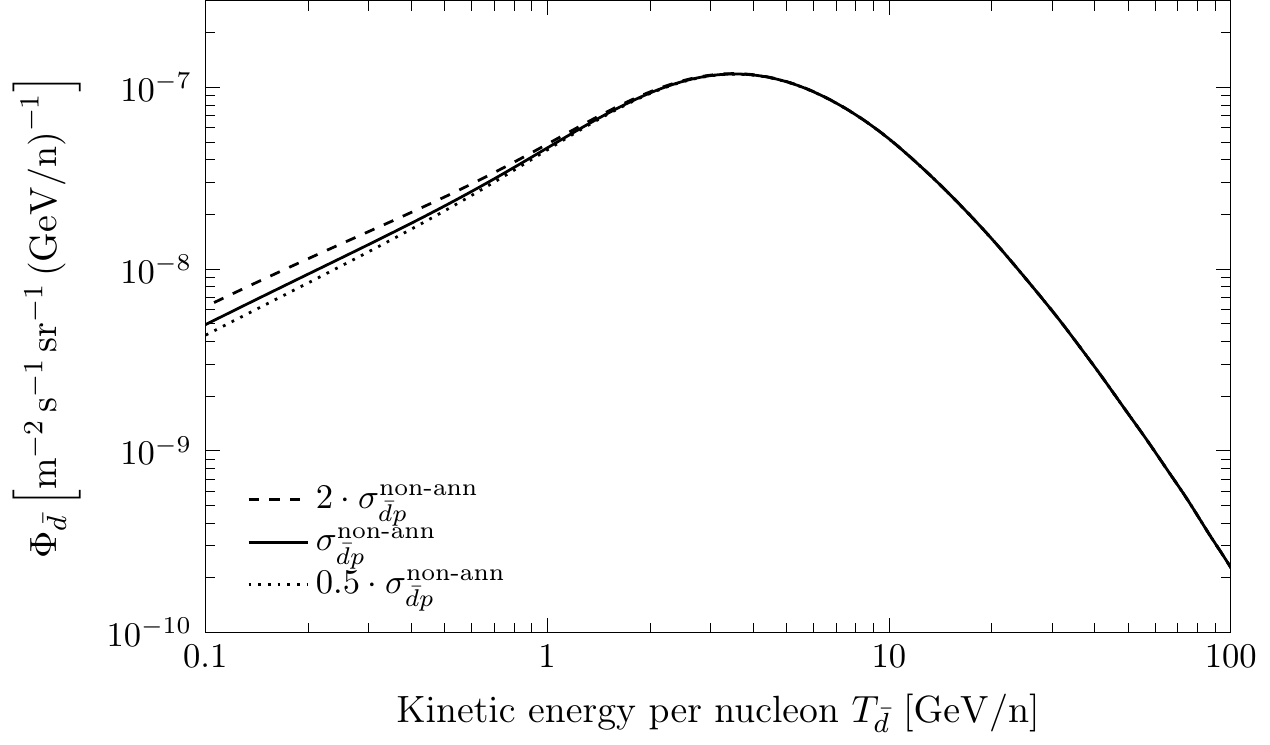}
\includegraphics[width=0.49\textwidth]{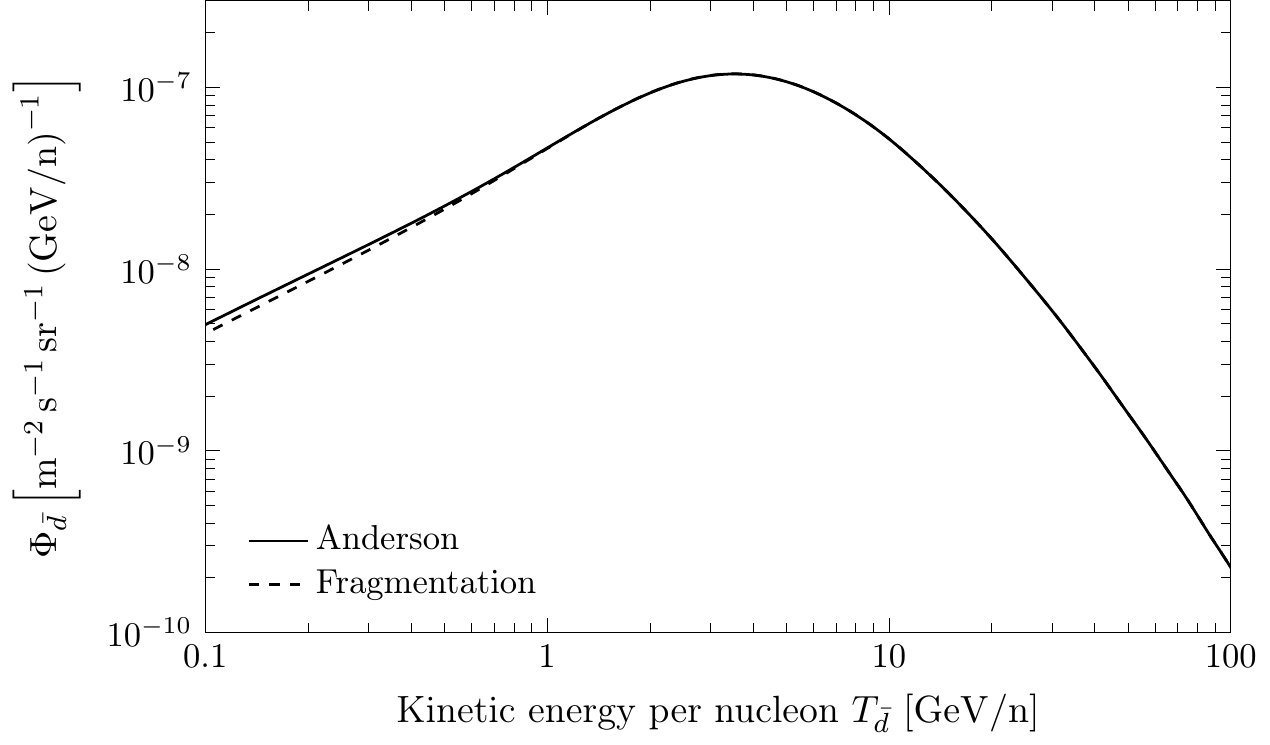}
\caption{Impact of various sources of uncertainty on the calculation of the cosmic antideuteron flux: value of the coalescence momentum, $p_0$ (upper left), choice of the propagation parameters (upper right), size of the non-annihilating antideuteron-proton rescattering cross section (lower left) and modeling of the non-annihilating rescattering cross section (lower right). See the text for details.}
\label{fig:uncertainties}
\end{center}
\end{figure}

The calculation of the cosmic antideuteron flux at the Earth suffers from a number of uncertainties. The largest uncertainty originates from the unknown value of the coalescence momentum $p_0$ in the interactions of cosmic ray nuclei with the interstellar gas. For our calculation we use $p_0 =152\,{\rm MeV}$, which is the value we extract from $p\,p$ collisions at $\sqrt{s}=53\,{\rm GeV}$ using the coalescence model implemented in the event-by-event generator. Nevertheless, as discussed in Section \ref{sec:source-term} (see also \cite{Ibarra:2012cc}) the coalescence momentum displays a dependence on the underlying process and on the center of mass energy involved. We then investigate in Fig.~\ref{fig:uncertainties}, upper left plot, the impact of varying the value of the coalescence momentum on the cosmic antideuteron flux, showing the result for $p_0= 133$ MeV, as extracted from the $\Upsilon$ decay, and for $p_0=192$ MeV, as extracted from the $Z$ decay~\cite{Ibarra:2012cc}. We would like to stress here that this source of uncertainty arises due to the very scarce experimental information on antideuteron production in $p\,p$ collision, which led us to use a Monte Carlo generator to simulate the production. This source of uncertainty could then be greatly reduced with a dedicated experiment measuring the differential antideuteron production cross section in $p\,p$ collisions at $\sqrt{s} \simeq 10\,{\rm GeV}$ (or $T_p \simeq 50$ GeV in a fixed target experiment), which is the most relevant process and center of mass energy for the calculation of the cosmic antideuteron flux, {\it cf.} Fig.~\ref{fig:Qsec_contris}.

The second largest uncertainty of our computation of the cosmic antideuteron flux stems from the degeneracies in the propagation parameters. We show in Fig.~\ref{fig:uncertainties}, upper right plot, the resulting flux for the three different sets of propagation parameters, MIN, MED and MAX, in Table \ref{tab:MinMedMax}. As apparent from the plot, the MIN and MED propagation parameters lead to nearly the same result, while the MAX parameters lead to a cosmic flux which is lower by $\sim 40$ \% in the low-energy region. The reason for this dependence is the following:  for kinetic energies above $\sim 5\,{\rm GeV/n}$, where energy loss effects are unimportant, the impact of the choice of propagation parameters is rather mild, due to the correlation between the production and propagation mechanisms of secondary antideuterons and of the secondary nuclei on which the propagation models are tuned. However, for small kinetic energies, the propagation history of antideuterons, which is dominated by the adiabatic energy loss, is very different to that of other secondary nuclei. This contribution is proportional to $V_c$ (see Eq.~(\ref{eqn:diff_eqn})), {\it i.e.} a larger value of the convection velocity implies a larger adiabatic energy loss. Therefore, the background flux is significantly lower for the MAX propagation parameters, as in this case the convection velocity $V_c = 5\,{\rm km/s}$ is rather low, while for the MED and MIN values this parameter is larger and nearly the same, with $V_c = 12\,{\rm km/s}$ ($V_c = 13.5\,{\rm km/s}$) for MED (MIN). This source of uncertainty, nonetheless, is expected to be reduced in the near future with the improved measurements by AMS-02 of the flux ratios of stable secondary-to-primary and unstable cosmic ray species, such as boron-to-carbon and beryllium-10-to-beryllium-9~\cite{Pato:2010ih}.

A further source of uncertainty stems from the modeling of the non-annihilating rescattering  cross sections, which enters the computation of the tertiary source term defined in Eq.~(\ref{eqn:tertiary_comp}). As explained above, there is scarce experimental information on the cross sections of the rescattering processes $\bar{d} + p/\text{He} \rightarrow \bar{d} + X$, therefore it is necessary to make several assumptions and estimations in their evaluation, making a definite error estimation very difficult. We investigate in Fig.~\ref{fig:uncertainties}, lower left plot, the impact on the result when doubling or halving our estimation for $\sigma_{\bar{d}p}^{\text{non-ann}}$, which amounts to a pretty conservative error of at most $\sim 30\%$. This rather mild dependence is connected to the fact that the dominant effect governing the low-energy part of the antideuteron background flux is the adiabatic energy loss and not the non-annihilating rescattering. Lastly, we compare the result using the method described in this section with that followed in~\cite{Bergstrom:1999jc,Donato:2001ms} based on the ``limiting fragmentation hypothesis''~\cite{Tan:1983de}, and which assumes that the differential cross section for the NAR process is a simple step function with respect to the incident kinetic energy of the antideuteron~\cite{Duperray:2005si,Donato:2008yx}: 
\begin{align}
\label{eqn:NAR_Frag}
\frac{\D \sigma^{\text{Frag}} \left( \bar{d}(T_{\bar{d}}') + p \rightarrow \bar{d}(T_{\bar{d}}) + X \right)}{\D T_{\bar{d}}} \equiv \frac{\sigma_{\text{tot}} \left( \bar{d}(T_{\bar{d}}') + p \rightarrow \bar{d} + X \right)}{T_{\bar{d}}'} \, \Theta(T_{\bar{d}}'-T_{\bar{d}}) \,.
\end{align}
The results are shown in Fig.~\ref{fig:uncertainties}, lower right plot, and are practically the same for both parametrizations. Again, this source of uncertainty could be better understood and reduced with dedicated experiments studying antideuteron-nucleus collisions.

\section{Conclusions}
\label{sec:conclusions}

In this work we have investigated the production of cosmic antideuterons in high energy collisions of cosmic rays with the interstellar matter. An accurate calculation of the secondary flux is important not only to assess the prospects to observe antideuterons at AMS-02 or GAPS, but also to determine the background fluxes to search for primary antideuterons from exotic sources. This flux has been calculated in the past using a ``modified factorized coalescence model'' which, due to the scarce experimental information on antideuteron production in high energy collisions, required a number of approximations and {\it ad hoc} assumptions. We have argued that some of the underlying assumptions of this approach are not validated by experimental data. Therefore, we have instead performed a full event-by-event simulation of the antideuteron production implementing the coalescence model in the Monte Carlo generator DPMJET-III, which was designed to simulate low-energetic hadron-hadron collisions, such as those relevant for experiments searching for cosmic antideuterons. Then, we used an appropriate diffusion-convection equation for charged cosmic rays in order to translate the secondary antideuteron source spectrum into the expected flux of background antideuterons at Earth. We find a secondary antideuteron flux approximately a factor of two smaller than the most recent analyses and which amounts to 0.13 background events at AMS-02 and 0.024 events at GAPS ULDB.  Lastly, we have also studied various sources of uncertainties affecting the computation of the antideuteron background flux and we have emphasized the importance of dedicated experiments analyzing antideuteron production in proton-proton collisions at $\sqrt{s}\sim 10$ GeV to reduce the uncertainties. 

\vspace{0.5cm}
\section*{Acknowledgements}
We are grateful to Stefan Roesler for providing us the code of the Monte Carlo generator DPMJET-III. This work was partially supported by the DFG cluster of excellence ``Origin and Structure of the Universe.''



\bibliographystyle{JHEP}
\bibliography{dbarback}

\end{document}